\documentclass[twocolumn,twoside]{revtex4}
\usepackage{graphicx}
\usepackage{fancyhdr}
\usepackage{caption}
\usepackage{subcaption}
\usepackage{amsmath}

\begin{document}

\title{Experimental review of Lepton Flavor Violation searches}

%

\author{M.~Hern\'andez~Villanueva\\
on behalf of the Belle~II collaboration}
\affiliation{Deutsches Elektronen–Synchrotron, 22607 Hamburg, Germany}

\begin{abstract}
A review of the experimental status in searches for charged Lepton Flavor Violation (LFV) is presented. 
Searches for LFV in decays of leptons, heavy mesons, and bosons are explored, with an emphasis on the experiments providing the best constraints in each sector. 
In addition, prospects for upper limits by current and upcoming experiments in the next decade are discussed.
\end{abstract}

\maketitle

\thispagestyle{fancy}


\section{Introduction}

In the Standard Model (SM), Lepton Flavor is a conserved quantity given the left-handed chirality of massless neutrinos. The experimental results in neutrino oscillation show that neutrinos change flavor, enabling Lepton Flavor Violation (LFV) as an established fact. In a minimum extension of the SM where neutrinos have a non-zero mass, charged LFV is enabled via neutrino oscillations but heavily suppressed by GIM mechanisms with branching ratios of $\sim10^{-50}$, making them unobservable in current experiments. Most extensions of the Standard Model enhance the branching ratio of LFV channels up to 10$^{-7}$, making any significant observation a clear signature of physics Beyond Standard Model (BSM). On the other hand, setting upper limits on the branching ratio of LFV translates into limits on the parameters associated with BSM models. 

In the literature, there are several reviews where LFV is discussed exhaustively from the theoretical and experimental point of view~\cite{Bernstein:2013hba, Cei:2014jtm, Calibbi:2017uvl, Ardu:2022sbt}. In this talk, an experimental overview is presented focusing on the strongest limits set to the date. In addition, prospects on limits to be set in future facilities are shown. 

\section{LFV in muons}

\subsection{$\mu^+ \to e^+ \gamma$}

The decay $\mu^+ \to e^+ \gamma$ was the first search of a LFV mode, even before the neutrino was discovered, testing an alternative hypothesis without neutrinos to the decay of a muon~\cite{Hincks:1948vr}. Since then, many experiments have performed searches for the signature of $\mu^+ \to e^+ \gamma$: a final state in the center-of-mass system of a back-to-back, monochromatic positron, and photon with an energy of 52.8 MeV each. The two main background sources are (i) the irreducible background $\mu^+ \to e^+ \gamma \nu_e \bar{\nu}_\mu$ when the neutrinos are low energetic, and (ii) the "accidental" background $\mu^+ \to e^+ \nu_e \bar{\nu}_\mu$ combined with a photon from elsewhere. 

\begin{figure}[h]
    \centering
    \includegraphics[width=0.48\textwidth]{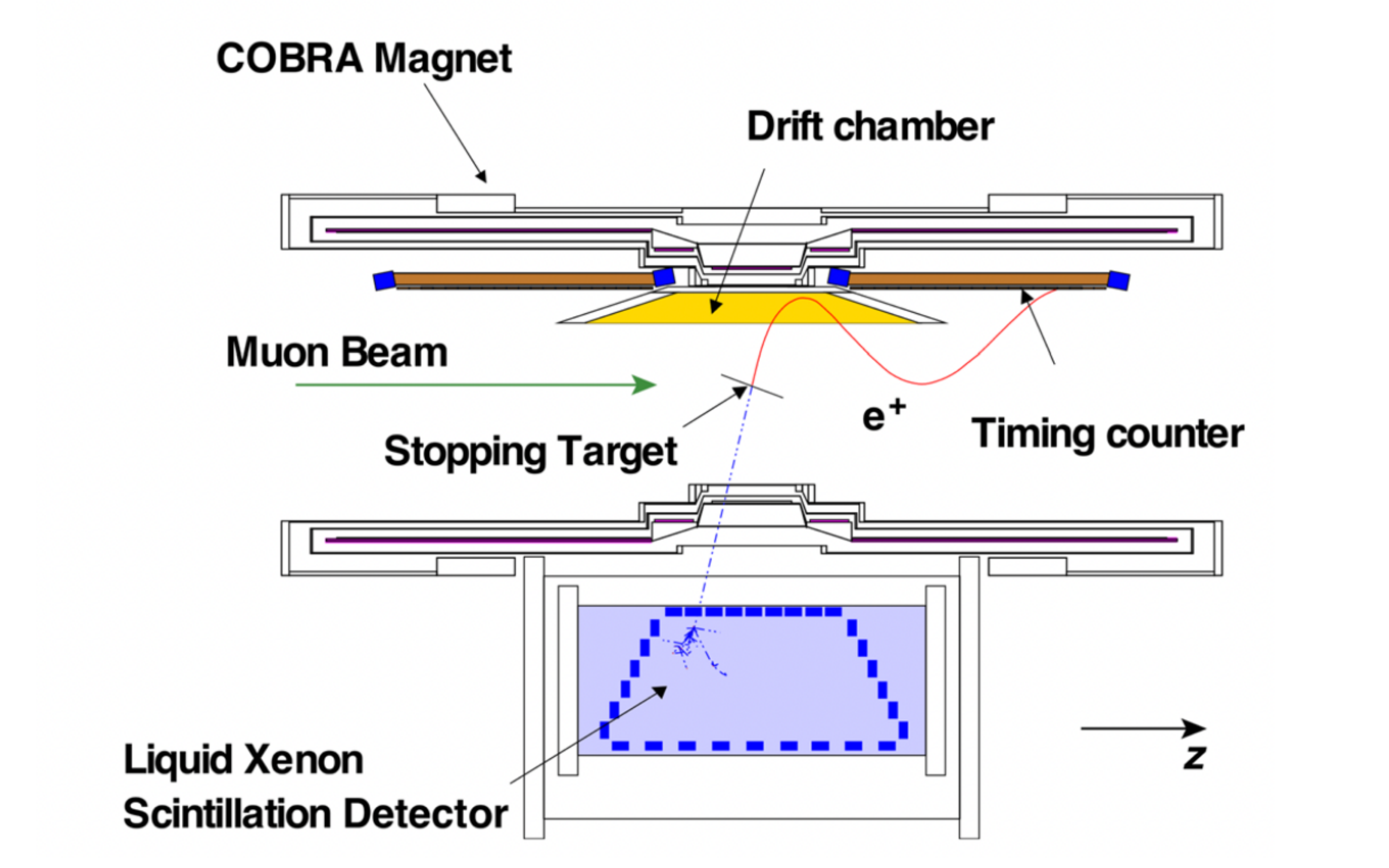}
\caption{Schematic view of the MEG experiment with the decay of a muon in the stopping target. Figure from~\cite{MEG:2016leq}.} 
\label{fig:MEGschema}
\end{figure}

The current best limit for $\mu^+ \to e^+ \gamma$ comes from the MEG experiment at the PSI laboratory. It consists of a target stopping muons from a $\mu^+$ beam, a drift chamber, scintillating timing counters, and a calorimeter immersed in the magnetic field generated by a superconducting magnet~\cite{Adam:2013vqa}. Figure~\ref{fig:MEGschema} shows the schematic view of the MEG experiment. 
With the full data set of $7.5\times10^{14}$ muons stopped on target from 2009 to 2013, MEG sets an upper limit for the branching ratio of ${\cal{B}}(\mu^+\to e^+\gamma) < 4.2\times 10^{-13}$ at 90\% C.L.~\cite{MEG:2016leq}. 

The upgraded experiment MEG-II at PSI is designed to record a muon decay rate twice that of the previous experiment, expecting to reach a sensitivity of $\sim 10^{-14}$ with 3 years of data taking~\cite{MEGII:2018kmf}. 

\subsection{$\mu^+ \to e^+ e^- e^+$}

The other possibility for searches of LFV in muon decays is via the channel $\mu^+ \to e^+ e^- e^+$, where two positrons and an electron coming from the same vertex are combined to search for an invariant mass matching the mass of the muon. Main backgrounds contributions come from the decays $\mu^+ \to e^+ e^- e^+  \nu_e \bar{\nu}_\mu$ when the neutrinos carry low energy, and accidental backgrounds from $\mu^+ \to e^+ \nu_e \bar{\nu}_\mu$ plus an electron-positron pair coming from photon conversion. 

The strongest limit for $\mu^+ \to e^+ e^- e^+$ comes from the SINDRUM experiment at PSI. A muon beam of 28 MeV/$c$ is stopped in the center of the detector at a rate of $\sim5\times10^6 \mu^+/s$, and the electrons and positrons coming from muons decays are detected in five concentric multiwire proportional chambers and a cylindrical array of 64 scintillation counters immersed in a magnetic field of 0.33 T. With no significant excess in the signal region, an upper limit of ${\cal{B}}(\mu^+\to e^+ e^- e^+) < 1.0\times 10^{-12}$ at 90\% C.L. is set~\cite{SINDRUM:1987nra}. 

\begin{figure}[h]
    \centering
    \includegraphics[width=0.44\textwidth]{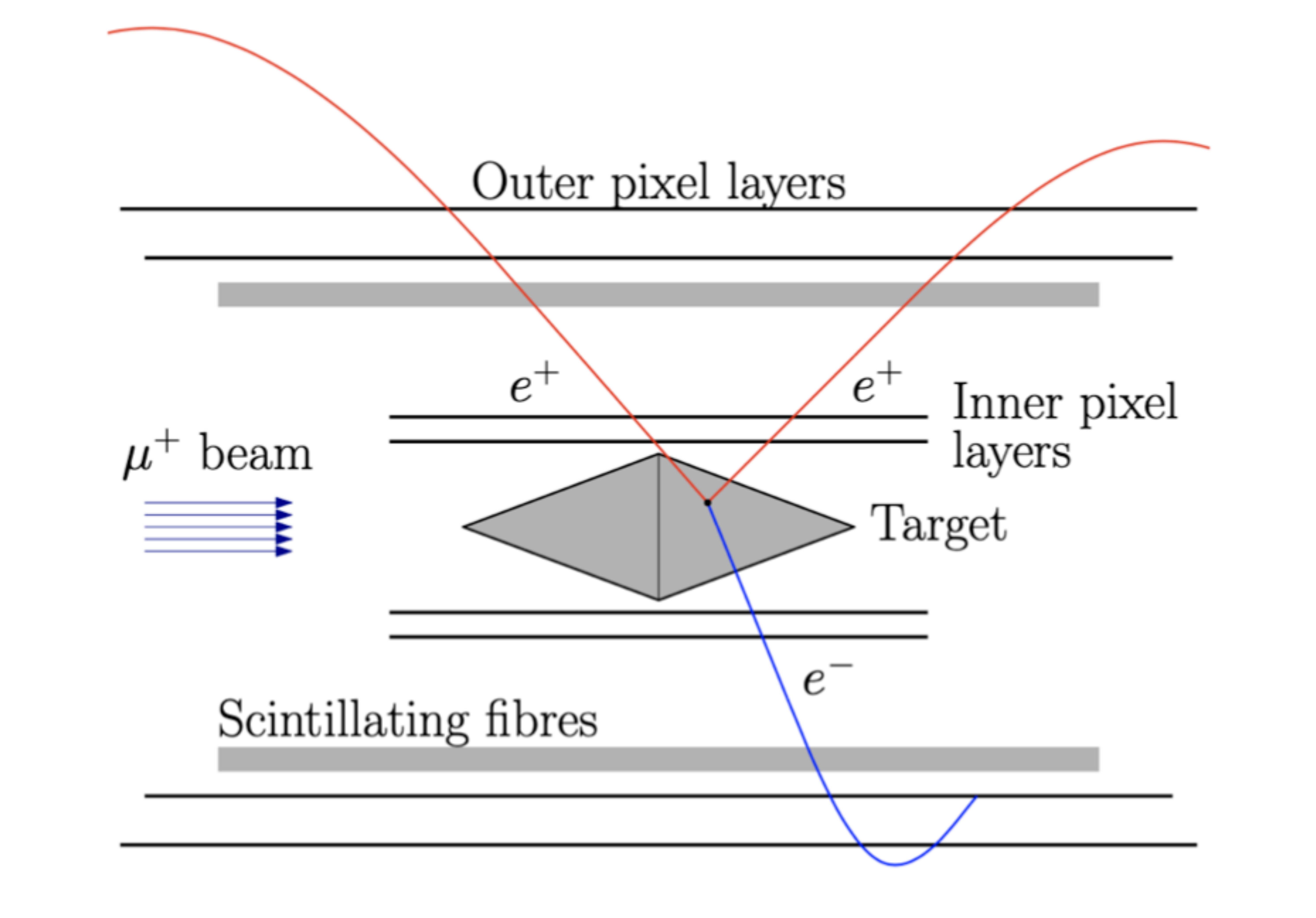}
\caption{Schematic view of the Mu3e experiment, simulating the decay of a muon in the stopping target. Figure from~\cite{Mu3e:2020gyw}.} 
\label{fig:Mu3eSchema}
\end{figure}

The Mu3e experiment at PSI aims for a sensitivity of $10^{-16}$ in the upper limit for the decay $\mu^+ \to e^+ e^- e^+$ by the end of the decade~\cite{Wauters:2021ghc}. Such an improvement of four orders of magnitude with respect to the current limit comes with experimental challenges to handle the projected rate of $10^9~\mu^+/s$, using low-density pixel detectors as a tracking system and a modern data acquisition system able to handle the high rates. Figure~\ref{fig:Mu3eSchema} shows the schematic view of the Mu3e experiment. 

\subsection{$\mu^- N \to e^- N$}

\begin{figure}[h]
    \centering
    \includegraphics[width=0.44\textwidth]{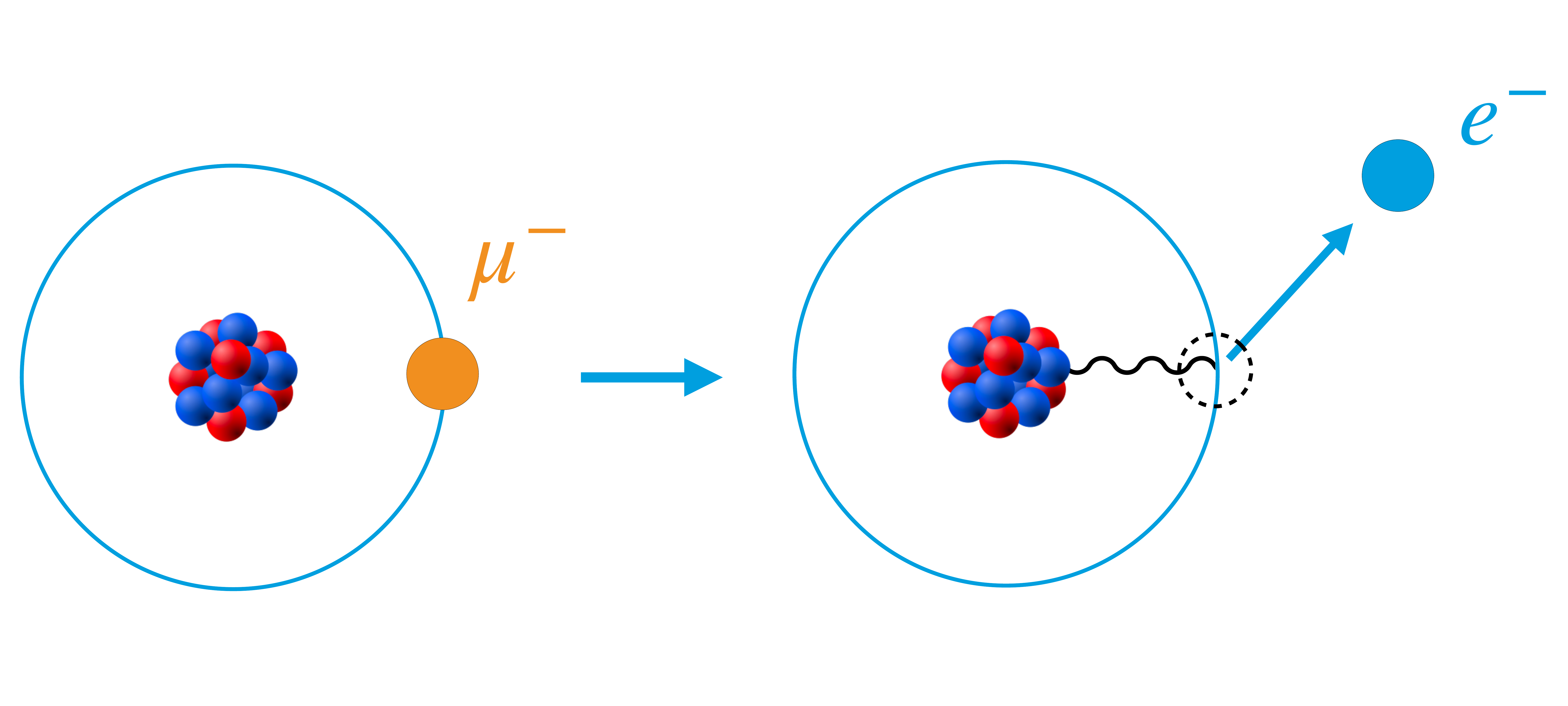}
\caption{$\mu-e$ conversion in the field of a nucleus.} 
\label{fig:mu2eNucleiDiagram}
\end{figure}

The conversion of a muon into an electron in the field of a nucleus represents a LFV process when no neutrinos are produced in the final state. In such a scenario, the kinematics of the process is modeled by a two-body decay as illustrated in Figure~\ref{fig:mu2eNucleiDiagram}, and the final state contains a mono-energetic electron with energy
\begin{equation}
    E_{\mu e} = m_\mu - E_{\rm binding} - E_{\rm recoil}, 
\end{equation}
where $m_\mu$ is the mass of the muon, $E_{\rm binding}$ is the binding energy of the 1s state, and $E_{\rm recoil}$ is the nuclear recoil energy for a muonic atom, dependent on the atomic nucleus~\cite{Lee:2022moh}. The main source of background is decay-in-orbit events where the captured muon decays to $\mu^- \to e^- \bar{\nu}_e \nu_\mu$, faking signal events when the energy of the electron is close to $E_{\mu e}$.

The current best limit comes from SINDRUM II at PSI, setting a limit in the rate of neutrinoless $\mu - e$ conversion relative to ordinary muon capture of
\begin{equation}
    \frac{\mu^- N \to e^- N}{\text{captured} \; \mu - N} < 3.3 \times 10^{-13}
\end{equation}
using a gold target that produces a mono-energetic electron with $E_{\mu e}=95.56$~MeV ~\cite{SINDRUMII:2006dvw}. 

Several collaborations are preparing experiments for improving the limits on $\mu^- N \to e^- N$ conversions. The Mu2e experiment under construction at Fermilab projects a sensitivity down to $O(10^{-17})$ using aluminum disks as stoping target that produce as signal electrons with $E_{\mu e} = 104.97$~MeV~\cite{Mu2e:2014fns}. The COMET experiment at J-PARC aims to reduce the upper limit down to $O(10^{-15})$ for Phase-I and $O(10^{-17})$ for Phase-II, using also aluminum as a target~\cite{COMET:2018auw}. The DeeMe experiment also at J-PARC is being prepared for a sensitivity of $O(10^{-13})$ using a SiC target~\cite{Natori:2014yba}. 


\section{LFV in tau leptons}

The $\tau$’s rate production at current colliders (10$^{9}$/yr) is much lower with respect to the production of muons in coming facilities (10$^{11}$/s). In addition, the lifetime of the $\tau$ lepton is significantly shorter and its decays must be studied in the production spot, without possibility of being transported like in the case of muons. 
However, thanks to the larger mass of the $\tau$ lepton, the BSM branching ratios can be orders of magnitude larger than the corresponding  muon decays. Moreover, the $\tau$ lepton is heavy enough to enable searches of neutrinoless semileptonic decays. 

Upper limits for LFV modes in $\tau$ lepton decays are mostly dominated by B-Factories  Belle~\cite{Belle:2000cnh} and BaBar~\cite{BaBar:2001yhh}, designed to collide with high-intensity electrons and positrons at the $\Upsilon(4S)$ energy in the center-of-mass system. The B-Factories share as common characteristics a well-defined initial state up to initial-state radiation (ISR), high vertex resolution, and dedicated subsystems for calorimetry and particle identification. The cross-section for production of $\tau$ lepton pairs is of the same order than for the production of B mesons, making the B-Factories optimal experiments for searches of rare $\tau$ lepton decays.

\subsection{$\tau^- \to \ell^- \gamma$}

\begin{figure}[h]
\centering
\begin{subfigure}[b]{0.24\textwidth}
    \centering
    \includegraphics[width=\textwidth]{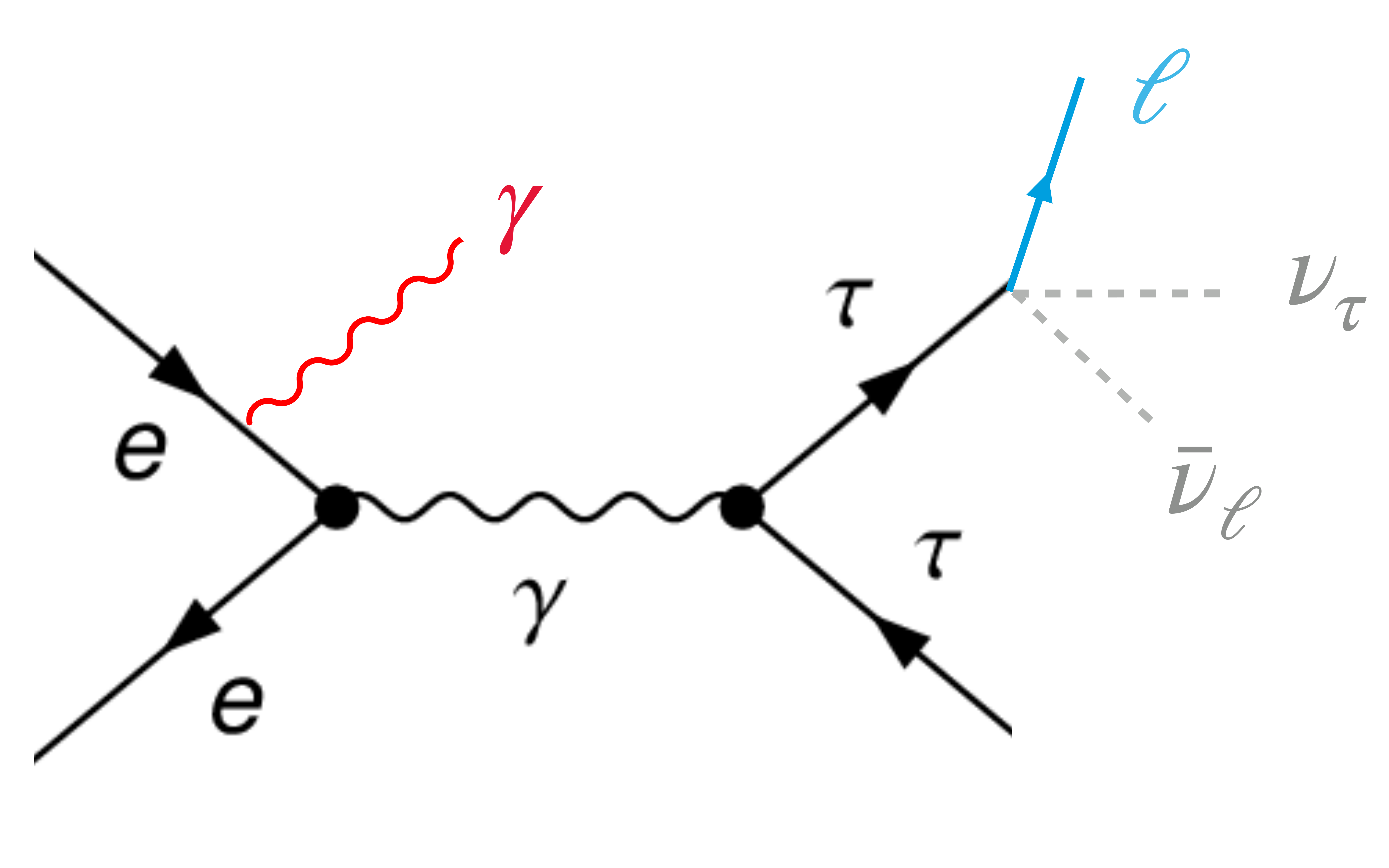}
    \caption{Irreducible background}
\end{subfigure}
\begin{subfigure}[b]{0.23\textwidth}
    \includegraphics[width=\textwidth]{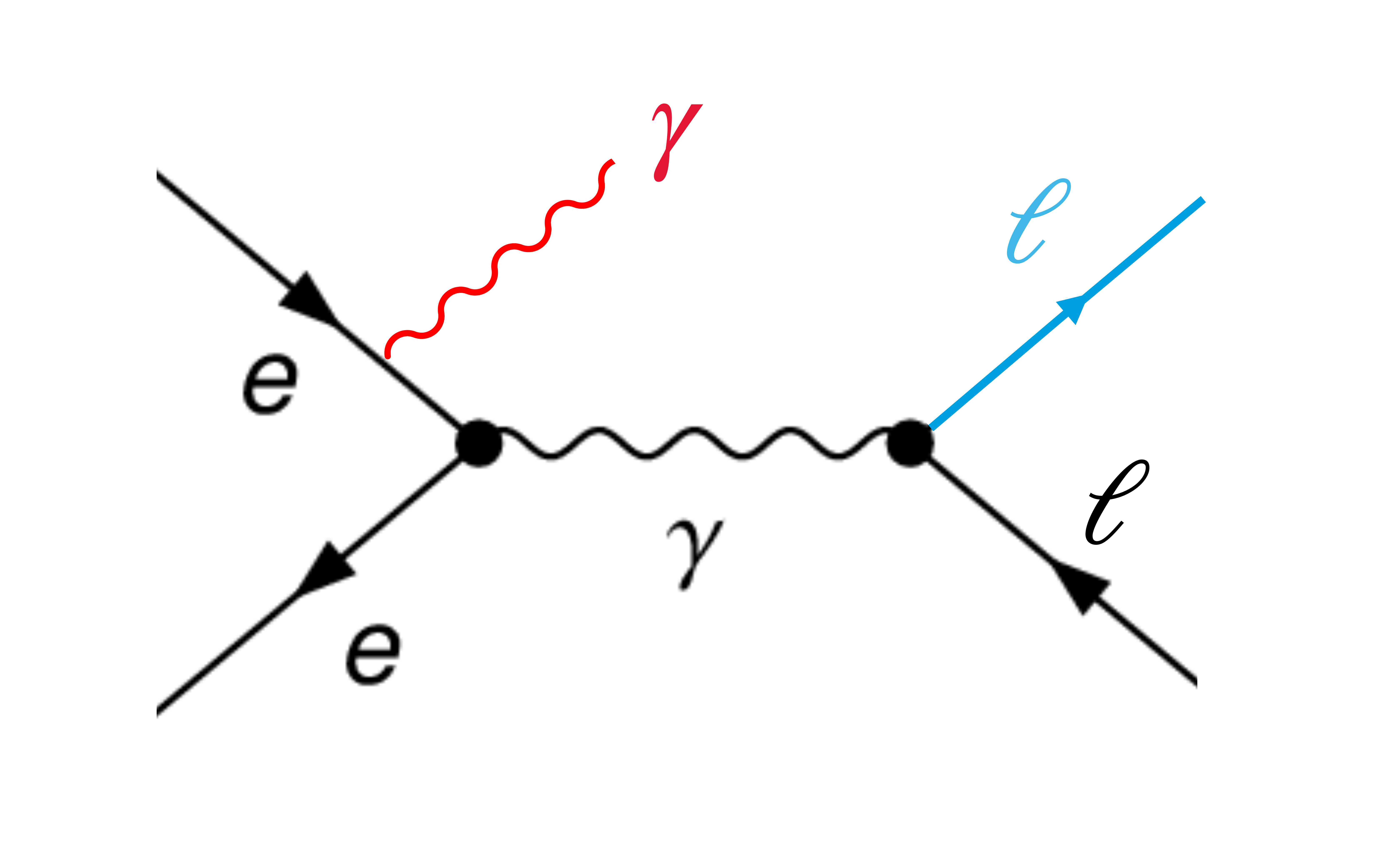}
    \caption{Mis-id tagging}
\end{subfigure}
\caption{Main background contributions in searches of the decay $\tau^+ \to \ell^+ \gamma$ in $e^+e^-$ collisions.} 
\label{fig:tau_ellgamma_bkgs}
\end{figure}

Even in presence of bounds for $\mu \to e \gamma$, rates for the counterparts $\tau^- \to \ell^- \gamma$ with $\ell^- = e^-, \mu^-$ can be higher by orders of magnitude.
Furthermore, $\tau^- \to \mu^- \gamma$ is considered one of the golden modes for the search of LFV in $\tau$ decays, having the largest branching ratio in models where the LFV is induced by heavy BSM particles in one-loop diagrams~\cite{Aushev:2010bq}. 
However, searches for $\tau^- \to \ell^- \gamma$ in $e^+e^-$ colliders are affected by strong backgrounds ilustrated in Figure~\ref{fig:tau_ellgamma_bkgs}. An irreducible background coming from $\tau^- \to \ell^- \bar{\nu}_\ell \nu_\tau$ decays plus a photon coming from elsewhere, and Bhabha or di-muon events from electron-positron collision misidentified as $\tau$ lepton pairs. 

\begin{figure}[h]
    \centering
    \includegraphics[width=0.48\textwidth]{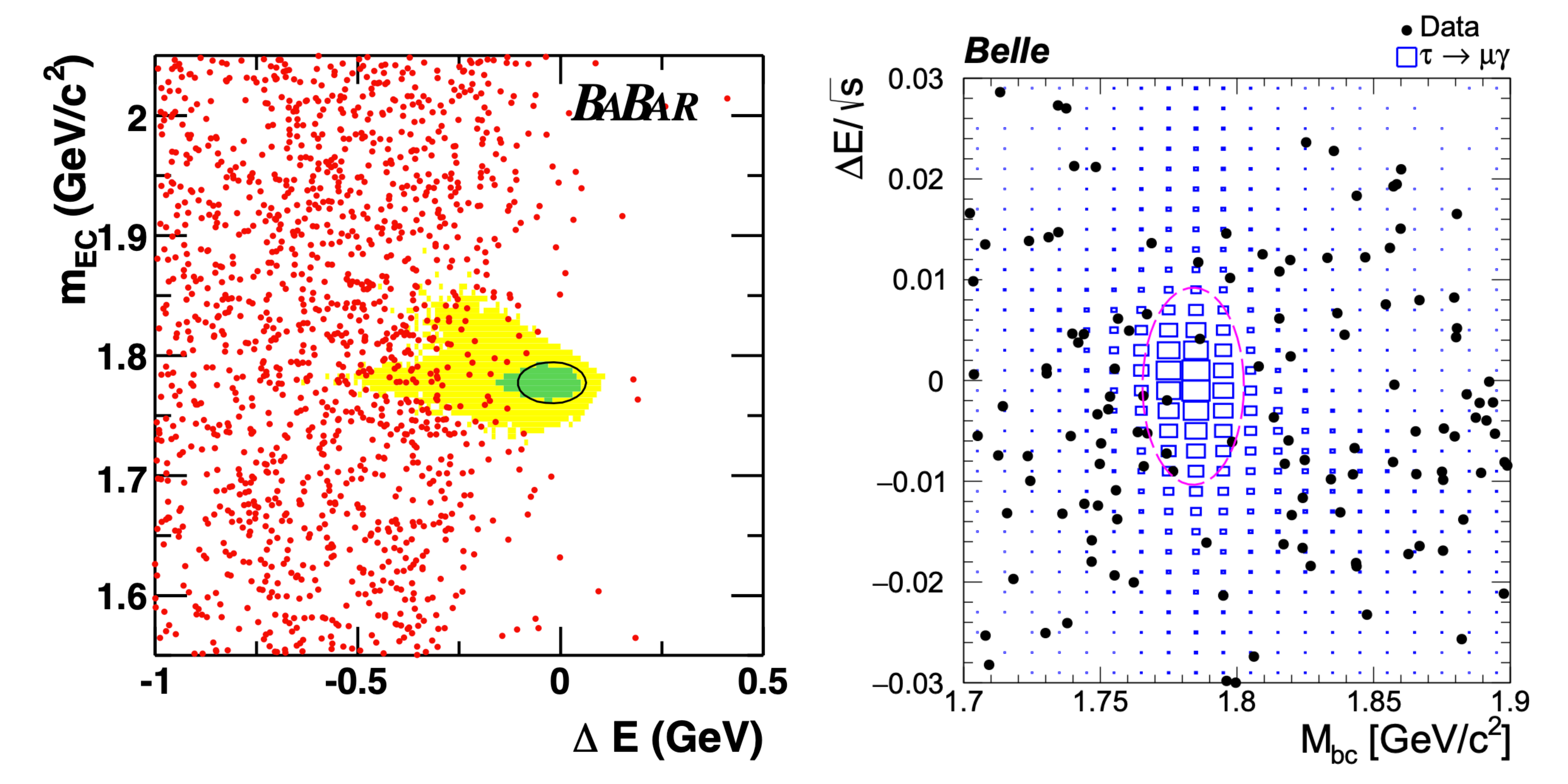}
\caption{Distribution of events in the 2D planes described in the text for $\tau \to e \gamma$ candidates at BaBar (left) and $\tau \to \mu \gamma$ candidates at Belle (right). The $2\sigma$ elliptical signal region is shown in both cases. Figures from~\cite{BaBar:2009hkt} and \cite{Belle:2021ysv}.} 
\label{fig:tau_ellgamma}
\end{figure}

Both B-Factories BaBar and Belle have searched for the LFV decays $\tau^- \to \ell^- \gamma$ with a recorded integrated luminosity of 516~fb$^{-1}$ and 988~fb$^{-1}$ respectively. 
Search for signal events is performed in a 2D region, where the kinematics of the 2-body decay establishes a well-defined signal region. 
BaBar identifies signal candidates with the beam-energy constrained $\tau$ mass ($m_{\rm EC}$), defined from a kinematic fit after requiring the center-of-mass $\tau$ energy to be $\sqrt{s}/2$ and assigning the origin of $\gamma$ to the point of closest approach of the signal lepton to the $e^+e^-$ axis, simultaneously with $\Delta E = (E^{CM}_{\ell \gamma} - \sqrt{s}/2)$; Belle uses the beam-constrained invariant mass ($M_{bc}$) vs $\Delta E$ phase space, being $M_{bc} = \sqrt{(E^{CM}_{\rm beam})^2 - |\vec{p}^{CM}_{\ell \gamma}|^2}$. Figure~\ref{fig:tau_ellgamma} shows remaining events after the selection criteria is applied for $\tau^- \to e^- \gamma$ and $\tau^- \to \mu^- \gamma$ candidates in BaBar and Belle respectively. As no significant excess has been found in the signal region~\cite{BaBar:2009hkt, Belle:2021ysv}, the best 90\% C.L. upper limits have been set as
\begin{equation}
\begin{aligned}
    {\cal{B}}(\tau^-\to e^- \gamma) &< 3.3\times 10^{-8} {\rm \hspace{0.5cm}(BaBar)},
    \\
    {\cal{B}}(\tau^-\to \mu^- \gamma) &< 4.2\times 10^{-8} {\rm \hspace{0.5cm}(Belle)}.
\end{aligned}
\end{equation}

The Belle~II experiment is the major upgrade and successor of Belle, expected to collect 50~ab$^{-1}$ by the end of its operation. The large statistics will allow a better understanding of the physics backgrounds for LFV modes. In addition, an increase in signal efficiency will be achieved at Belle II thanks to higher trigger efficiencies and improvements in the reconstruction algorithms. On the other hand, beam backgrounds are a potentially more serious concern than in the predecessor Belle.  The sensitivity expected at Belle~II for the $\tau \to \ell \gamma$ modes is at the $O(10^{-9})$ level with the full data set~\cite{Belle-II:2018jsg}. 

\subsection{$\tau^- \to \ell^- \ell^+ \ell^-$}

\begin{figure}[h]
    \centering
    \includegraphics[width=0.5\textwidth,trim={2cm 0 0 0},clip]{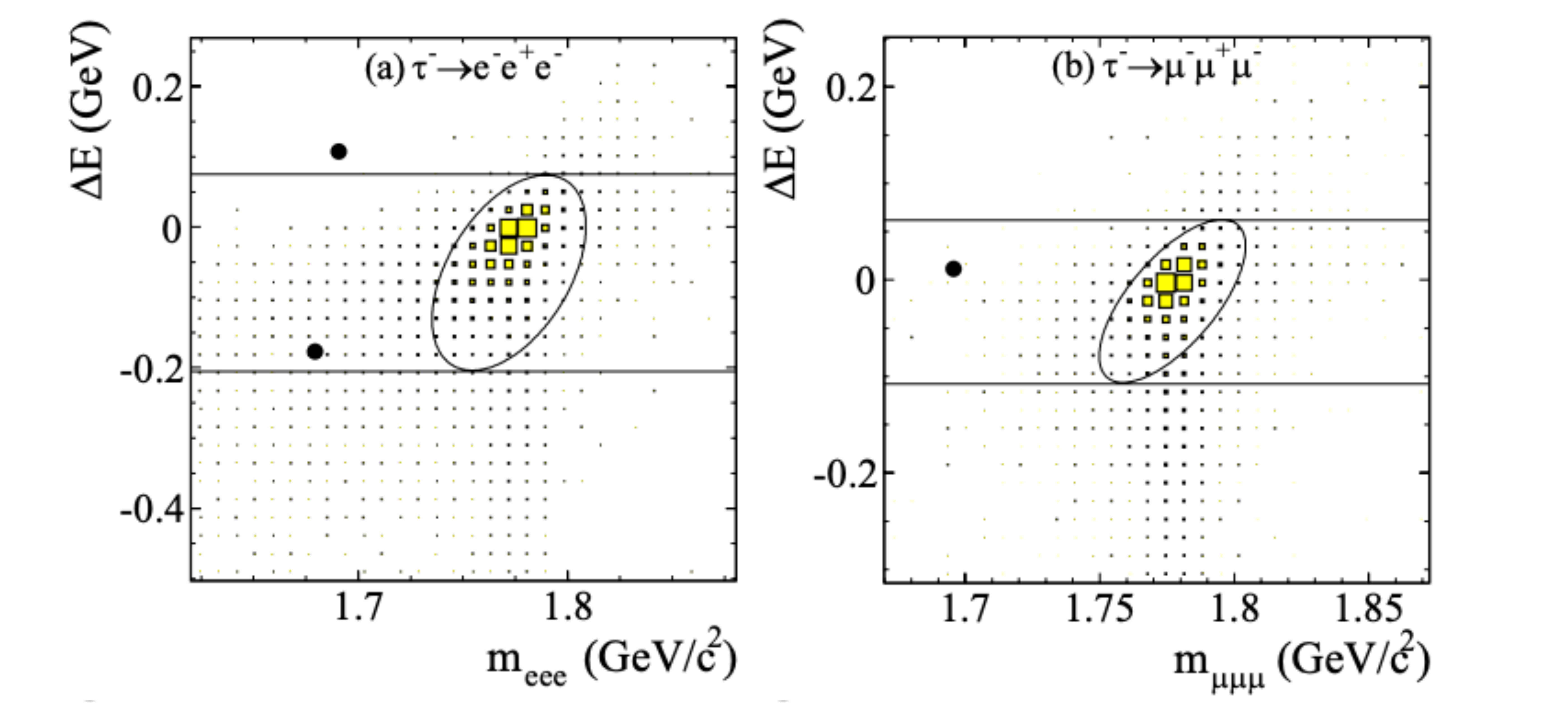}
\caption{Distribution of events in the 2D $m_{\ell \ell \ell}$ vs $\Delta E$ planes for $\tau^- \to e^- e^+ e^-$ (left) and $\tau^- \to \mu^- \mu^+ \mu^-$ (right) candidates at Belle. The elliptical signal region is shown in both cases. Figure from~\cite{Hayasaka:2010np}.} 
\label{fig:tau_3leptons}
\end{figure}

Belle and BaBar have also searched for the LFV decay $\tau^- \to \ell^- \ell^+ \ell^-$ in six possible combinations when $\ell = e \; {\rm or} \; \mu$, with Belle imposing the strongest upper limits~\cite{Hayasaka:2010np} listed at Table~\ref{tab:tau_3leptons}. 
Signal events are defined by four reconstructed tracks, combining 3 charged lepton candidates with invariant mass $m_{\ell \ell \ell}$ around the mass of the $\tau$ lepton and with $\Delta E$ around 0. 
An advantage of the 3 lepton LFV decay channels with respect to $\tau \to \ell \gamma$ modes is they are "background-free", in the sense that no background events are expected in the signal region. 
To illustrate, Figure~\ref{fig:tau_3leptons} shows remaining events in the $m_{\ell \ell \ell}$ vs $\Delta E$ at Belle for $\tau^- \to e^- e^+ e^-$ and $\tau^- \to \mu^- \mu^+ \mu^-$ after selection criteria is applied. It is clear the low background when compared to the plots shown for $\tau \to \ell \gamma$ and the absence of events close to the signal region.

\begin{table}[h]
    \centering
    \caption{Current best upper limits in the branching ratios of LFV $\tau$ decays to three leptons, imposed by the Belle experiment~\cite{Hayasaka:2010np}.}
    \begin{tabular}{|l|c|c|}
    \hline
      \textbf{Mode} & \textbf{\; U.L. (90\% C.L.) \;}  \\ \hline
      $\tau^- \to e^- e^+e^-$ \; \;& $2.7 \times 10^{-8}$  \\ \hline
      $\tau^- \to \mu^- \mu^+\mu^-$ \; \; & $2.1 \times 10^{-8}$  \\ \hline
      $\tau^- \to \mu^- e^+e^-$ \; \; & $1.8 \times 10^{-8}$  \\ \hline
      $\tau^- \to \mu^- \mu^+e^-$ \; \; & $2.7 \times 10^{-8}$ \\ \hline
      $\tau^- \to e^+ \mu^-\mu^-$ \; \; & $1.7 \times 10^{-8}$  \\ \hline
      $\tau^- \to \mu^+ e^- e^-$ \; \; & $1.5 \times 10^{-8}$ \\ \hline
    \end{tabular}
    \label{tab:tau_3leptons}
\end{table}

Of particular interest is the decay $\tau^- \to \mu^-\mu^+\mu^-$, which can be searched in proton-proton collisions at the LHC experiments thanks to a clean signature of muons in the trigger systems. Searches are performed via $W^-$ or heavy hadron decays containing a $\tau$ lepton with the subsequent LFV decay. The strongest limit from proton collisions is set by the LHCb experiment, searching $\tau^- \to \mu^-\mu^+\mu^-$ events in $\tau$ leptons coming from semileptonic decays of b and c hadrons. Signal events are reconstructed from 3 muon candidates, and two multi-variable analysis classifiers are used to perform signal discrimination. Without a significant excess in the defined signal region, the 90\% C.L. upper limit set by LHCb is ${\cal{B}}(\tau^- \to \mu^- \mu^+\mu^-)~<~4.6~\times~10^{-8}$~\cite{LHCb:2014kws}.

\subsection{Semileptonic $\tau$ LFV modes}

\begin{figure*}[t]
\centering
\includegraphics[width=145mm]{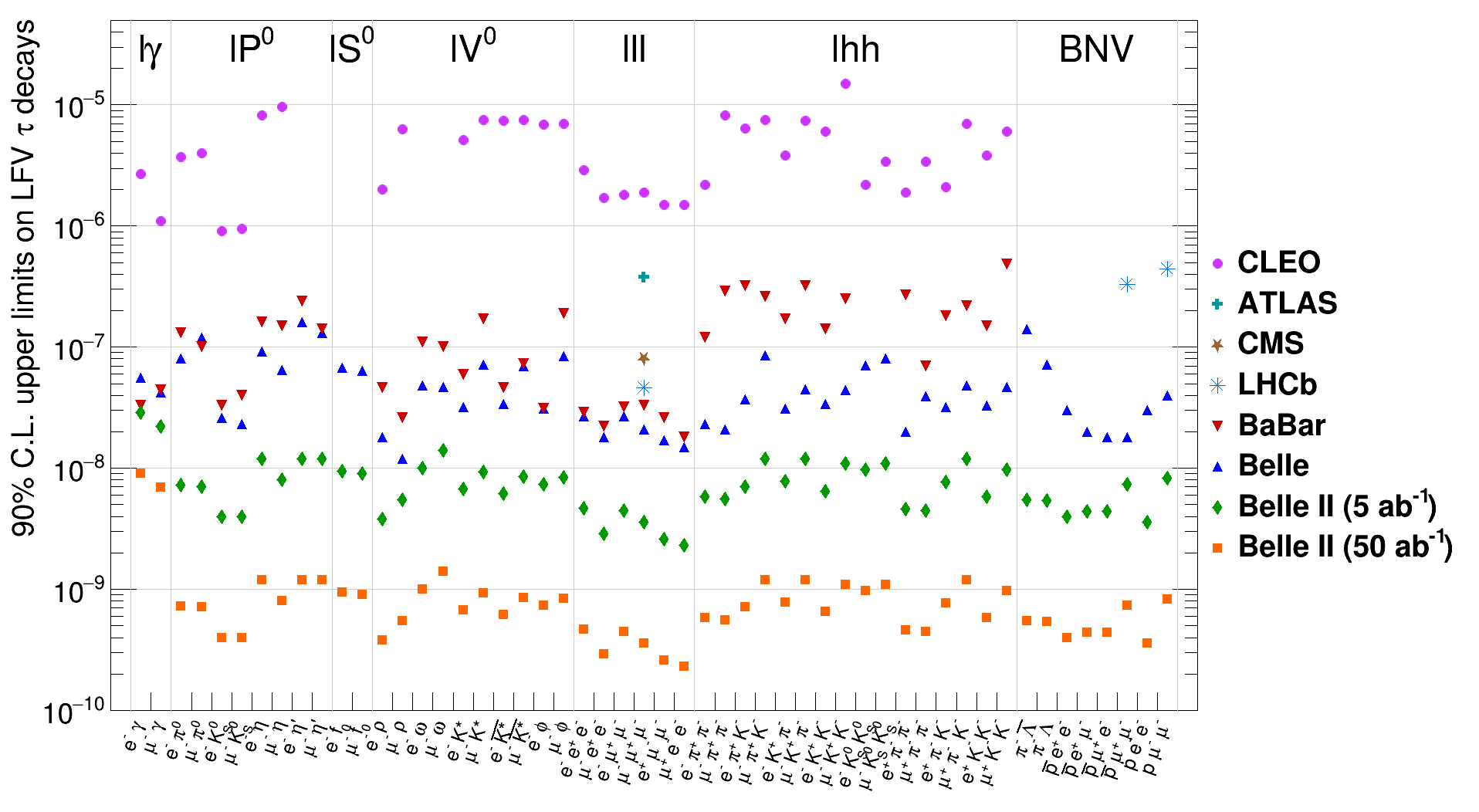}
\caption{Current observed upper limits at CLEO, BaBar, Belle, ATLAS, CMS and LHCb experiments for the 52 LFV 2-body and 3-body $\tau$ lepton decays, as well as projections for expected upper limits at Belle~II. Figure from~\cite{Banerjee:2022xuw}.} \label{fig:TauLFVmodes}
\end{figure*}

The $\tau$ is the only lepton heavy enough to decay into hadrons, enabling semileptonic LFV modes. For the searches of $\tau$ LFV modes, semileptonic decays are classified as a lepton and a pseudoscalar meson: $\tau^-\to \ell^- P^0$ with $S^0 = \pi^0, \eta, \eta^\prime, K_S^0$; a lepton and a scalar meson $\tau^- \to \ell^- f_0(980)$; a lepton and a neutral vector meson: $\tau^- \to \ell^- V^0$ with $V^0 = \rho, \omega, K^*(892), \phi$; or a lepton and two mesons $\tau^- \to \ell^- h_1 h_2$.
Figure~\ref{fig:TauLFVmodes} show the 52 neutrinoless 2-body and 3-body decay modes searched by CLEO~\cite{CLEO:1982pvq}, the B-Factories and the LHC experiments, as well as projections for the upper limits to be set at Belle~II with 5 and 50~ab$^{-1}$ of integrated luminosity recorded.

\section{Searches of LFV with B mesons}

Recent hints of lepton non-universality observed in semileptonic decays of B mesons and $b \to s \ell \ell$ transitions~\cite{Crivellin:2021sff, Ferreres-Sole:2022qrs} could be explained by BSM scenarios that enable LFV in B meson decays, such as models with leptoquarks~\cite{Becirevic:2016yqi} or a new $Z'$ boson~\cite{Crivellin:2015era}.

\subsection{$B^0_{(s)} \to e \mu$ decays}

In scenarios with BSM interactions, the branching ratio for $B^0_{(s)} \to e^\pm \mu^\mp$ decays can be enhanced up to O($10^{-11})$.
Best upper limits for $B^0_{(s)} \to e^\pm \mu^\mp$ are imposed by the LHCb experiment. Search is performed with 3~fb$^{-1}$ of data collected from $pp$ collisions at 7 and 8 TeV.
Signal candidates are identified by combining two tracks identified as a muon and an electron, with a common vertex and invariant mass around the mass of the $B^0_{(s)}$ meson. Major background contributions come from B decays with misidentified or missing final states.
A  boosted decision tree (BDT) is used to discriminate signal events, using both candidates with and without bremsstrahlung correction.
Without significant excess of events in the signal region~\cite{LHCb:2017hag}, the 90\% C.L. upper limits set by LHCb are
\begin{equation}
\begin{aligned}
    {\cal{B}}(B^0 \to e^\pm \mu^\mp) &< 1.3\times 10^{-9},
    \\
    {\cal{B}}(B_s^0 \to e^\pm \mu^\mp) &< 6.3\times 10^{-9}.
\end{aligned}
\end{equation}

\subsection{$B^0_{(s)} \to \tau \ell$ decays}

Another possibility is to explore two-lepton LFV decays of the B mesons with a $\tau$ lepton included. Final states involving a $\tau$ make the reconstruction harder due to the presence of missing energy in the signal candidate.  

The LHCb experiment has set the strongest limits for the branching ratio of $B^0_{(s)} \to \tau^\pm\mu^\mp$. With $\tau$ lepton candidates reconstructed from $\tau^- \to \pi^-\pi^+\pi^-\nu_\tau$, a simultaneous fit on 4 regions of the response of a BDT is used to search for an excess of events around the mass of the $B$ meson. Assuming no cross-feed contamination between both FLV channels given the overlap between the $B^0$ and $B^0_s$ signal regions, the upper limits
\begin{equation}
\begin{aligned}
    {\cal{B}}(B^0 \to \tau^\pm\mu^\mp) &< 1.4\times 10^{-5},
    \\
    {\cal{B}}(B_s^0 \to \tau^\pm\mu^\mp) &< 4.2\times 10^{-5},
\end{aligned}
\end{equation}
at 95\% C.L. are obtained~\cite{LHCb:2019ujz}. 

In the case of the LFV decay to a $\tau$ lepton and an electron, the Belle experiment imposes the strongest constrain for the branching ratio of $B^0 \to \tau^\pm e^\mp$. Since the initial state for the $B$-pair is well defined, a hadronic B tag is used to infer the momentum of the B meson on the opposite side and the $\tau$ lepton candidate does not require to be reconstructed. Instead, the missing invariant mass of the event is determined and signal candidates are searched around the mass of the $\tau$ lepton. Without an excess with respect to the background level expectation, Belle sets an upper limit at 90\% C.L. of ${\cal{B}}(B^0 \to \tau^\pm e^\mp) < 1.6 \times 10^{-5}$~\cite{Belle:2021rod}.

\subsection{$b \to s \ell \ell'$ transitions}

Extensions of the SM that enable LFV $b \to s \ell \ell'$ transitions predict branching ratios up to the range $10^8-10^{10}$. Table~\ref{tab:bslltransitions} lists best limits to the date for the transitions with the three generations of leptons involved, showing less stringent limits when $\tau$ leptons are included in the final state due to the missing energy in the reconstruction process. 
Models of LFV can produce signatures with different charge configurations, and both limits are provided in addition to the sum.

Upgrades I and II at LHCb, as well as searches with the full data set to be collected by Belle II, project an improvement in the upper limits of one order of magnitude for the next decade. Early results with the full-event interpretation algorithm at Belle II show higher tag-side reconstruction efficiency~\cite{Belle-II:2021rof, Dattola:2021cmw}, opening the possibility for further constraints in the limits with inclusive tagging techniques.


\begin{table}[h]
    \centering
    \caption{Limits in branching ratios for $b \to s \ell \ell^\prime$ transitions.}
    \begin{tabular}{|l|c|c|}
    \hline
      \textbf{Mode} & \textbf{U.L. (90\% C.L.)} & \textbf{Experiment} \\ \hline
      $B^+ \to K^+ \mu^- e^+$ & $7.0 \times 10^{-9}$	& LHCb~\cite{LHCb:2019bix} \\ \hline
	 $B^+ \to K^+ \mu^+ e^-$ & $6.4 \times 10^{-9}$ & LHCb~\cite{LHCb:2019bix}\\  \hline
	 $B^0 \to K^0 \mu^\pm e^\mp$& $ 3.8 \times 10^{-8}$	& Belle~\cite{BELLE:2019xld} \\ \hline
	 $B^0 \to K^{*0} \mu^+ e^-$& $ 5.7 \times 10^{-9}$	& LHCb~\cite{LHCb:2022lrd} \\ \hline
	 $B^0 \to K^{*0} \mu^- e^+$& $ 6.7 \times 10^{-9}$	& LHCb~\cite{LHCb:2022lrd} \\ \hline
	 $B^0 \to K^{*0} \mu^\pm e^\mp$& $ 9.9 \times 10^{-9}$	& LHCb~\cite{LHCb:2022lrd} \\ \hline
	 $B^0 \to \phi \mu^\pm e^\mp$& $ 1.6 \times 10^{-8}$	& LHCb~\cite{LHCb:2022lrd} \\ \hline
	 $B^+ \to K^+ \tau^\pm \mu^\mp$& $ 4.8 \times 10^{-5}$	& BaBar~\cite{BaBar:2012azg} \\ \hline
	 $B^+ \to K^+ \tau^\pm e^\mp$& $ 3.0 \times 10^{-5}$	& BaBar~\cite{BaBar:2012azg} \\ \hline
    \end{tabular}
    \label{tab:bslltransitions}
\end{table}

\section{LFV searches with bosons}

Searches of LFV modes in the decays of the $Z$ and Higgs bosons are accessible only at high-energy colliders and complementary to LFV searches in leptons and mesons.
On the other hand, any BSM mechanism that produces LFV interactions in bosons would enable them also in low-energy processes. In consequence, limits from LFV decays of muons and tau leptons set indirect limits on the maximum branching ratios for $Z/H \to \ell \ell'$ decays~\cite{Calibbi:2021pyh}.

\subsection{$Z^0 \to \ell \ell^\prime$ decays}

\begin{figure}[h]
    \centering
    \includegraphics[width=0.4\textwidth]{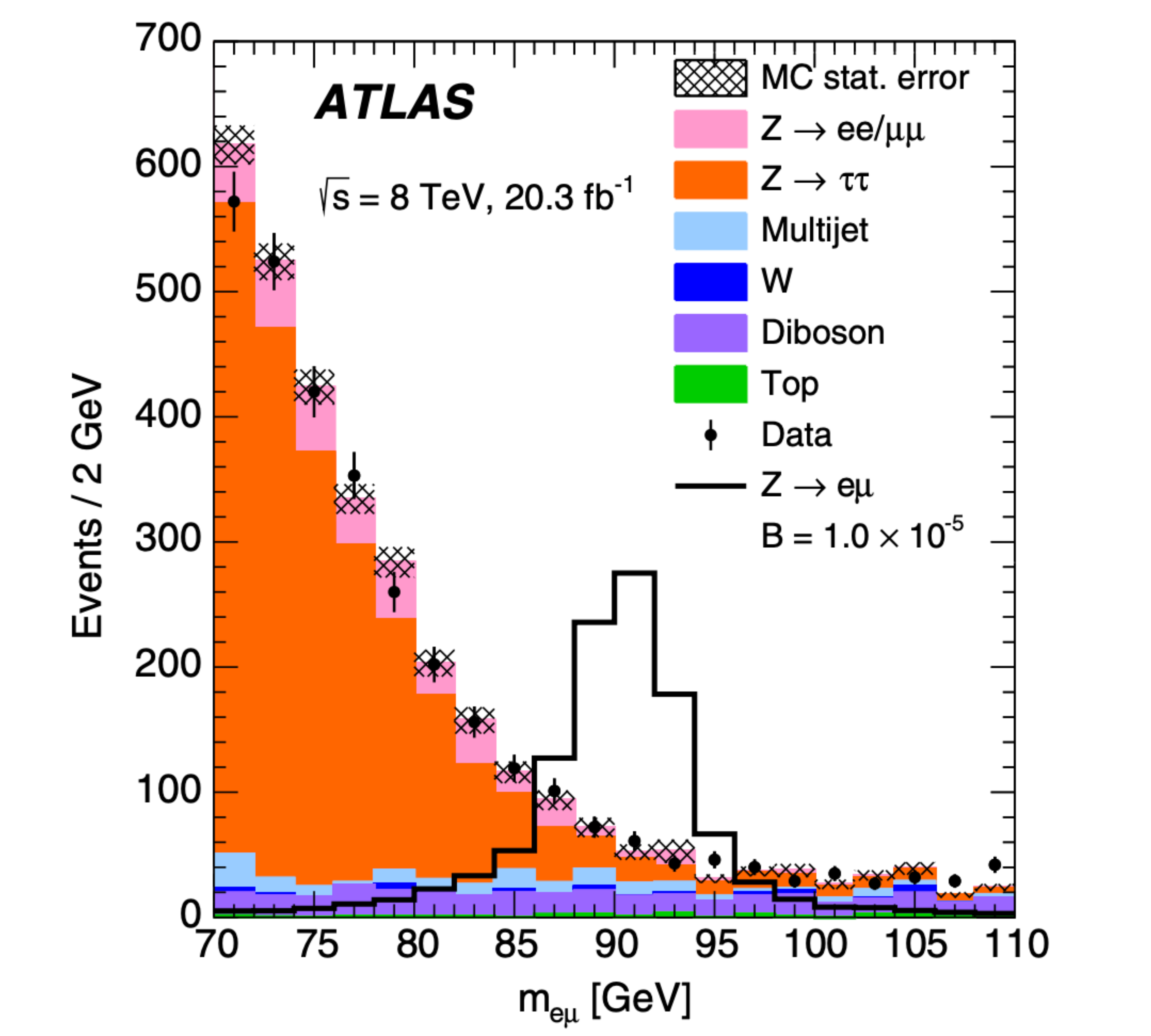}
\caption{Invariant mass distribution of $Z^0 \to e\mu$ candidates in data collected by ATLAS, with background expectation from MC processes after the selection cuts are applied. Figure from~\cite{ATLAS:2014vur}.} 
\label{fig:Z_emu_Atlas}
\end{figure}

In the case of the LFV mode with leptons of first and second generation only, reconstruction of $Z^0 \to e^\pm \mu^\mp$ candidates is performed from tracks associated with an electron and a muon, and searching for an excess in the invariant mass $m_{e\mu}$ of the signal candidates around the mass of the $Z^0$ boson. The ATLAS experiment has set the best limit for $Z \to e^\pm\mu^\mp$ using 20.3 fb$^{-1}$ of data collected in $pp$ collisions at 8 TeV. Main background comes from $Z^0 \to \tau^+\tau^-$ with the subsequent leptonic decay the $\tau$ lepton, as illustrated by Figure~\ref{fig:Z_emu_Atlas}. The mass spectrum is consistent with the background expectations, and an upper limit of ${\cal{B}}(Z^0 \to e^\pm \mu^\mp) < 7.5\times 10^{-7}$ at 95\% C.L. is set~\cite{ATLAS:2014vur}.

Limits with a $\tau$ lepton in the final state are still dominated by the LEP experiments OPAL and DELPHI. The reconstruction of $Z^0 \to \tau^\pm \ell^\mp$ decays with $\ell = e, \mu$ is performed with criteria optimized for the identification of an electron or a muon and a charged cone associated with a $\tau$ decay. A cone consists of charged tracks and electromagnetic clusters within a cone of half-angle 35$^\circ$. The background events are composed mostly of $Z^0 \to \tau^+ \tau^-$ events with subsequent $\tau^- \to \ell^- \bar{\nu}_\ell \nu_\tau$ in one of the leptons. The observed candidates are consistent with the expected background, and limits at the 95\% C.L. are set~\cite{OPAL:1995grn, DELPHI:1996iox}:
\begin{equation}
\begin{aligned}
    {\cal{B}}(Z^0 \to \tau^\pm e^\mp) &< 9.8\times 10^{-6} {\rm \hspace{0.5cm}(OPAL)},
    \\
    {\cal{B}}(Z^0 \to \tau^\pm \mu^\mp) &< 1.2\times 10^{-5} {\rm \hspace{0.5cm}(DELPHI)}.
\end{aligned}
\end{equation}

In all $Z^0$ LFV modes, an improvement in the upper limits by a factor of 10 is expected with the data collected by the high-luminosity LHC. 

\subsection{$H \to \ell \ell^\prime$ decays}
 
The best limit for the LFV mode for a Higgs boson $H \to e^\pm \mu^\mp$ is imposed by the ATLAS experiment, using 139 fb$^{-1}$ of data collected in $pp$ collisions at 13 TeV.  The search is performed with a similar strategy as described in the previous section, combining an electron and a muon candidates and performing a fit in the invariant mass $m_{e\mu}$ distribution.
Events from top quarks are suppressed by identifying b-hadrons in the final state, using a multivariate algorithm with calorimeter and tracking information, and the remaining background events come from $Z\to\tau\tau$ and W + jets. 
Without a significant excess in the signal, an upper limit of ${\cal{B}}(H \to e^\pm \mu^\mp) < 6.1\times 10^{-7}$ at 95\% C.L. is established~\cite{ATLAS:2019old}.

The most stringent limits for $H \to \tau^\pm e^\mp$ and $H \to \tau^\pm e^\mp$ are also provided by ATLAS, based on a data set of $pp$ collisions at 13 TeV corresponding to 36~fb$^{-1}$. Both leptonic and hadronic decays of the $\tau$ lepton are exploited, with the lepton from the Higgs and the $\tau$ of a different flavor to reject the strong di-lepton background from Dell-Yan processes. The analysis exploits boosted decision tree algorithms for the separation of signal from background events, being $Z \to \tau^+\tau^-$ and pairs of top quarks the most significant background components. 
In the absence of a significant excess, the upper limits 
\begin{equation}
\begin{aligned}
    {\cal{B}}(H \to \tau^\pm e^\mp) &< 4.7\times 10^{-3},
    \\
    {\cal{B}}(H \to \tau^\pm \mu^\mp) &< 2.5\times 10^{-3},
\end{aligned}
\end{equation}
at 95\% C.L. are set for a Higgs boson with $m_H = 125~$GeV/c$^{2}$~\cite{ATLAS:2019pmk}. The upper limits are computed assuming no cross-feed contamination between the channels.

\section{Summary}

A brief experimental overview of LFV searches has been presented, with emphasis on the strongest limits set to the date.
References have been provided for the reader, where the details of each result can be consulted. 

Other possibilities not discussed during this talk, but not less relevant, include searches of LFV in decays of kaons: $K_L^0 \to \mu^\pm e^\mp$~\cite{BNL:1998apv}, $K^+ \to \pi^+ \mu^+ e^-$~\cite{Appel:2000wg}; $J/\psi$ decays: $J/\psi \to \mu^\pm e^\mp$~\cite{BESIII:2022exh}, $J/\psi \to \tau^\pm e^\mp$~\cite{BESIII:2021slj}, $J/\psi \to \tau^\mp \mu^\mp$~\cite{BES:2004jiw}; decays with BSM particles involved: $\ell^- \to \ell'^- \alpha$~\cite{ARGUS:1995bjh, Ibarra:2021xyk}, $Z' \to \ell \ell'$~\cite{Belle-II:2019qfb}, and several more not covered due to time limitations.

In all the sectors presented, an improvement of 1-2 orders of magnitude is expected with the data collected at the future muon experiment facilities, Belle~II at KEK, and the upgrades of ATLAS, CMS, and LHCb.
The experiments in the next decade will not only rely on larger statistics but also on improved hardware and tools that will increase the signal efficiency, and in a better understanding of the background contributions.

\bibliography{references}

\end{document}